# Cartographie du stress thermique au sein d'une cours d'école parisienne : Couplage de mesures microclimatiques fixes et mobiles

**KARAM G. (1,2), CHANIAL M. (1,3), CHAUMONT M. (1), HENDEL M. (1,2), et ROYON L. (1)**

*(1) Université de Paris, LIED, UMR 8236, CNRS, F-75013, Paris, France, ghid.karam@esiee.fr*
*(2) Université Gustave Eiffel, ESIEE Paris, Département SEN, F-93162, Noisy-le-Grand, France*
*(3) Mairie de Paris, Direction de la Propreté et de l'Eau & de la Voirie et des Déplacements, Paris, France*

**Résumé :** *Le dérèglement climatique sera à l'origine de vagues de chaleurs plus fréquentes, plus intenses et de plus longue durée à l'horizon 2050. Dans le cadre de son Plan Climat et de sa stratégie de résilience, la Ville de Paris déploie, à travers son programme Oasis, un maillage d'îlots de fraicheur urbains. Des travaux de réaménagements sont conduits au sein des cours d'écoles afin de réduire le stress thermique des usagers. Sur la base d'un cas d'étude, nous établissons une première méthodologie afin de quantifier l'impact microclimatique de la transformation d'une cour d'école par des mesures microclimatiques mobiles lors de vagues de chaleur. Celles-ci sont couplées avec des données de stations fixes situées hors de la cour pour quantifier l'évolution du stress thermique, évalué par l'UTCI. La cartographie du stress thermique ainsi obtenue permet un premier diagnostic microclimatique de la cour d'école.*

**Mots clés :** *Mesures microclimatiques mobiles, UTCI, cartographie, couplage.*

**Summary:** *Climate change will result in more frequent, more intense and longer-lasting heat waves by 2050. As part of its Climate Plan and its resilience strategy, the City of Paris is deploying, through its Oasis program, a network of urban cool islands to mitigate the urban heat island phenomena: schoolyards are renovated in order to reduce the heat stress of users. We establish a methodology aiming to quantify the microclimatic impact of the transformation. Mobile measurements are carried out within a case courtyard under hot conditions and coupled with fixed weather station data to evaluate heat stress using UTCI. The heat stress mapping thus obtained allows a first microclimatic diagnosis of the schoolyard.*

## Introduction

Outre l'augmentation des températures annuelles moyennes, le dérèglement climatique sera à l'origine de vagues de chaleurs plus fréquentes, plus intenses et de plus longue durée à l'horizon 2050 (Meehl and Tebaldi 2004). L'adaptation des villes à ces nouveaux extrêmes fait l'objet de recherches visant à mettre en place des parcours et îlots de fraicheur urbains afin d'atténuer les causes et effets de ces phénomènes (Lai et al. 2019).

Dans le cadre de son plan Climat et de sa Stratégie de Résilience (Mairie de Paris 2017, 2018), la Ville de Paris prévoit de déployer, à travers son programme Oasis, un maillage d'îlots de fraicheur urbains constitué de cours d'écoles débitumées et végétalisées. Le maillage de cours d'écoles ainsi transformées permettra de pallier le manque d'espaces verts de la capitale et s'inscrit dans une stratégie de rafraîchissement urbain à la fois globale et de proximité.

L'évaluation *in situ* de la performance rafraîchissante des travaux est un enjeu important pour la Ville de Paris. L'étude de l'impact microclimatique de techniques de rafraîchissement urbain est le sujet de nombreux travaux expérimentaux et numériques menés à différentes échelles du territoire. Parmi les travaux expérimentaux, certains s'appuient sur des mesures mobiles, d'autres fixes (Mohegh et al. 2018; Hendel 2016). De manière plus générale, l'observation de l'évolution des sites avant et après application des stratégies de végétalisation est nécessaire à l'évaluation de l'efficacité de ces dernières (Bowler et al. 2010).

Nous nous concentrons dans cet article sur les mesures mobiles qui permettent d'établir une cartographie du stress thermique avant et après les travaux de réaménagement. Une synthèse tenant compte de diverses corrections temporelles permettra de diagnostiquer de manière quantitative l'impact microclimatique des travaux de réaménagement au sein de la cour de l'Ecole Daumesnil. La méthode proposée ici sera mise en œuvre dans le cadre du projet FEDER UIA OASIS pour l'évaluation du rafraîchissement obtenu dans 10 cours d'école pilotes.

1. **Matériels et méthode**

   *1.1 Description de la cour d'école*

La cour d'école étudiée est située au 70 avenue Daumesnil dans le 12ᵉ arrondissement de Paris. Des travaux ont été réalisés dans la cour à l'été 2018 et ont consisté à remplacer le revêtement en asphalte par un revêtement drainant et réfléchissant, ajouter de la terre végétale au pied des arbres existants et planter deux nouveaux arbres. La façade du bâtiment situé au Nord entièrement vitrée a de même été équipée de protections solaires. La Figure 1 donne une vue aérienne du site avant et après travaux.

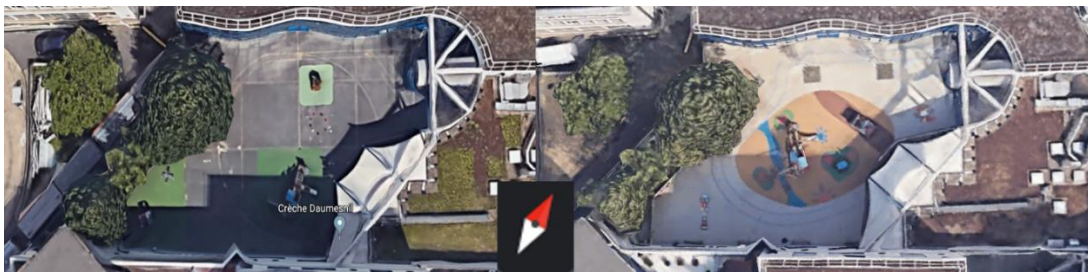

Figure 1 : Vue aérienne de la cour Daumesnil avant (gauche) et après rénovation (droite)

L'albédo du revêtement avant et après travaux a été mesuré selon la méthode ASTM E1918-16 (2016) : l'asphalte noir et le sol amortissant vert des espaces de jeux ont un albédo de 0,11, tandis que le nouveau revêtement beige a un albédo de 0,3.

   *1.2 Stations microclimatiques*

Afin d'évaluer l'impact de la transformation, des mesures ont été effectuées en été avant (juin 2018) et après (juin 2019) travaux, durant des journées chaudes et radiatives, c'est-à-dire de classe de stabilité de Pasquill A ou A-B (Pasquill 1961) : vent faible et ciel dégagé (couverture ≤ 3 octas), températures maximales et minimales supérieures à 25°C et à 16°C, respectivement.

Tableau 1 : Récapitulatif des sondes équipant la station microclimatique mobile et leur incertitude.

|  | *Paramètre* | *Hauteur* | *Incertitude* |
|---|---|---|---|
| *Globe Noir* | $T_g$ | 1,5 m | 1°C |
| *Anémomètre à fil chaud* | $v_a$ | 1,5 m | 0.3m/s |
| *Thermohygromètre sous abri* | $T_{air}$<br>$RH$ | 1,5 m | 0,5°C<br>2.5% |

L'instrumentation de la station de mesure mobile utilisée (DeltaOhm HD32.1) est récapitulée au Tableau 1. Les mesures sont effectuées à 1,5 m de hauteur toutes les 15 secondes. La Figure 2 montre les différents points de mesure au sein de la cour. Les points sont choisis de sorte à être représentatifs des différents environnements de la cour : plein soleil, ombre, proximité de végétation. En revanche, ils ne permettent pas de caractériser l'impact de la plantation des deux arbres dont la position n'était pas connue au moment de définir les points de mesure. Chaque point est caractérisé pendant 10 à 20 minutes afin de permettre au globe noir de se stabiliser, selon qu'il y ait ou non un changement de conditions d'ensoleillement (ombragé ou ensoleillé). L'intégralité des mesures est effectuée entre 13h45 et 15h. les condition d'ensoleillement des points 4 et 5 ayant changé d'une année à l'autre, (M4 et M5 à l'ombre en 2018 et au soleil en 2019), nous les avons retirés de l'analyse.

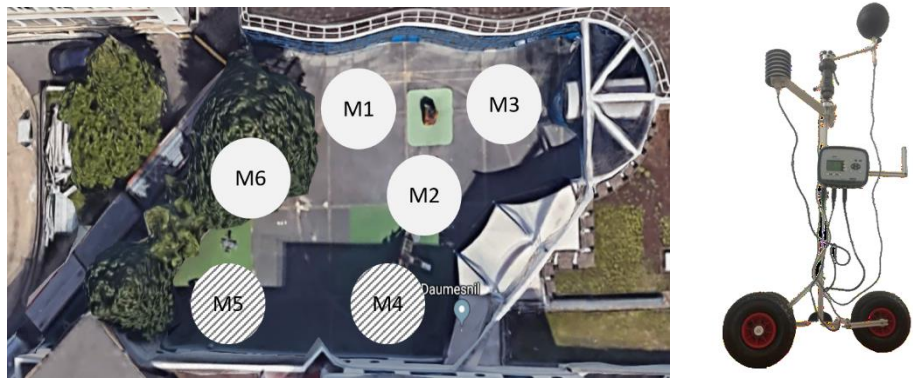

Figure 2 : Points de mesures dans la cour (gauche) et photo de la station mobile (droite).

Ces mesures sont utilisées pour calculer l'UTCI afin de quantifier le stress thermique ressenti par un usager (adulte) de la cour (Błazejczyk et al. 2010).

### 1.3 Méthode d'analyse

Il est établi qu'une correction temporelle est nécessaire pour des données obtenues au cours de traversées mobiles réalisées sur plusieurs heures afin de prendre en compte l'évolution des paramètres mesurés au cours de la période de mesure (Tsin et al. 2016; Leconte et al. 2015). De même, les mesures microclimatiques ayant lieu à une année d'intervalle, il est nécessaire de tenir compte des variations synoptiques entre ces deux traversées. L'appariement de ces mesures avec une station fixe et servant de témoin permettra de prendre en compte les variations temporelles à petites et grandes échelles.

Afin de quantifier l'impact de la transformation de la cour, on s'intéresse à l'indicateur
$$\Delta T_{UTCI,i} = T_{UTCI,i}^{mobile} - T_{UTCI,i}^{ref}(T_{air,i}^{témoin}, RH_i^{témoin}, T_{mrt,ref}, v_{ref})$$

Où $\Delta T_{UTCI,i}$ indique en un point de mesure i, l'écart entre la température équivalente UTCI calculée en ce point ($T_{UTCI,i}$) et une température équivalente UTCI de référence $T_{UTCI,i}^{ref}$

La température équivalente UTCI de référence correspond à l'état de stress thermique d'un piéton situé dans une cour ombragée à l'abri des courants d'air. Cette dernière est calculée en utilisant la température de l'air et l'humidité relative mesurées à la station témoin. On impose $T_{mrt,ref} = T_{air,i}^{témoin}$ en on fixe la vitesse de vent à $0.5\ m/s$ à 1,5 m de hauteur.

Cela permet de s'affranchir des variations de conditions météorologiques entre les années ainsi que de l'évolution temporelle des paramètres au cours d'une même campagne de mesure.

Avec cet indicateur, on ne s'intéresse non pas à un niveau absolu de stress thermique – qui dépend nécessairement des conditions météorologiques synoptiques et pas seulement de la performance intrinsèque du site étudié – mais à l'écart de stress par rapport au niveau de référence défini. Cet écart est attribué a priori aux modifications du site d'étude. Nous proposons une nouvelle échelle inspirée de l'échelle de l'UTCI qui permette de rendre compte de l'ampleur du changement vis-à-vis de l'échelle UTCI. Ce dernier permet de mettre en évidence un écart faible (de 0 à 3°C), modéré (de 3 à 6°) qui peut ne pas impacter la classification de stress thermique. Un écart conséquent (6 à 9°C) ou important (supérieur à 9°C) fait basculer la catégorie de stress thermique d'une ou de deux catégories.

La station météo témoin la plus proche dont nous disposons est située dans le 14e arrondissement de Paris, près du métro Alésia. Les données y sont disponibles à la minute.

## 2. Résultats et Discussion

Les mesures ont été réalisées le 20 juin 2018 avant et le 24 juin 2019 après travaux entre 13h45 et 15h. Les températures minimales et maximales enregistrées à la station météorologique du parc Montsouris ont été de 16,6 °C et 29,1 °C, respectivement le 20 juin 2018 pour une vitesse moyenne du vent de 2 m/s. Le 24 juin 2019, ces valeurs sont de 21°C et 34°C pour une vitesse moyenne du vent de 2,7 m/s.

Le Tableau 2 résume les observations de l'écart $\Delta T_{UTCI}$ aux 6 points de mesure de la cour. La Figure 3 illustre la répartition spatiale des $\Delta UTCI$ calculés au sein de la cour avant (gauche) et après (droite) travaux.

Tableau 2 : $\Delta T_{UTCI}$ relevés au sein de la cour avant (juin 2018) et après (juin 2019) travaux (en °C)

|  | M1 | M2 | M3 | M6 |
|---|---|---|---|---|
| *20 juin 2018* | +10,2 | +9,9 | +8,8 | +1 |
| *24 juin 2019* | +10,9 | +11,7 | +9,7 | +3,1 |
| $\Delta T_{UTCI,2019} - \Delta T_{UTCI,2018}$ | +0,7 | +1.8 | +0.9 | +3 |

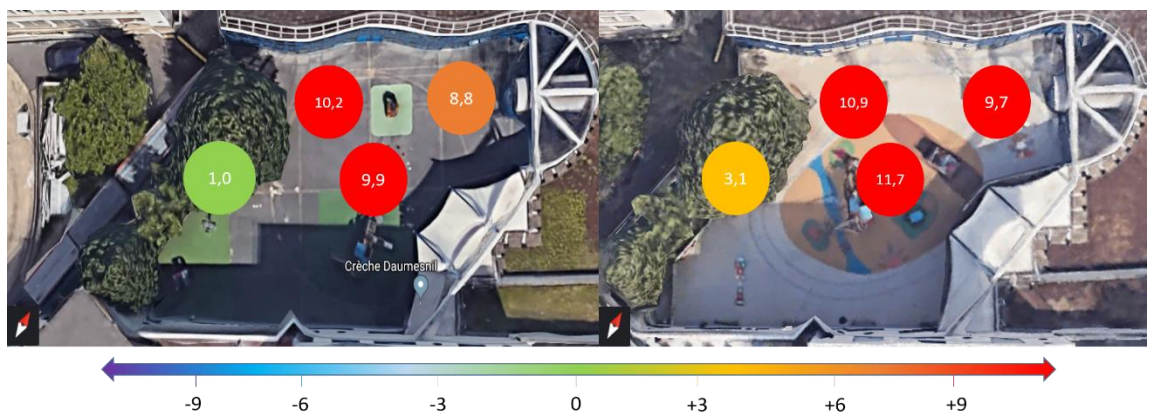

Figure 3 : Cartographie du surstress thermique au sein de la cour. Gauche : avant travaux ; droite : après travaux. En bas, code couleur des écarts UTCI.

En plein soleil, l'écart des températures UTCI par rapport aux températures de référence est de l'ordre de la dizaine de degrés Celsius et présente un stress thermique absolu de niveau élevé. Il est maintenu voire augmenté pour le point M3 d'une année à l'autre (augmentation de l'ordre de 1°C). L'écart $\Delta T_{UTCI}$ relevé sous l'arbre augmente de façon marginale de 2°C, faisant basculer ce point d'une catégorie de stress neutre à celle d'un stress modéré, bien que ce point soit situé sous l'arbre et que l'ancien enrobé qui s'y trouvait a été remplacé par de la terre végétale. L'amplitude des variations de $\Delta T_{UTCI}$ d'une année à l'autre aux points ensoleillés étant de l'ordre de 1°C, une étude de la propagation des incertitudes dues aux instruments de mesures et leurs répercussion sur l'UTCI permettrait de nuancer les observations.

Ces mesures tendent à indiquer que le bilan radiatif dégradé d'un piéton dans la cour compense les effets rafraîchissant souvent attendus du remplacement des revêtements par des matériaux à albédo plus important. L'augmentation de $T_{mrt}$ sous l'arbre a notamment amplifié une situation où l'écart $\Delta T_{air}$ reste très peu marqué ($\Delta T_{air}$ passe de +0,78°C en 2018 à +1,66°C en 2019). Une pluviométrie 15% plus faible en 2019 que 2018 ayant pu contribuer à un stress hydrique de la végétation peut aussi figurer parmi les facteurs de la dégradation de ce stress thermique.

L'observation des écarts de température moyenne radiante observées en chaque point par rapport à la température de référence montre une diminution de ces dernières en plein soleil, mais une augmentation de ces dernières sous l'arbre (Figure 4). Ces observations peuvent être corrélées avec l'augmentation d'albédo qui va rafraichir les surfaces ensoleillées, mais contrebalancer l'évapotranspiration et l'ombrage sous l'arbre (Erell et al. 2014). Toutefois cette diminution semble contrée par une augmentation du $\Delta T_{air}$ de l'ordre de 3°C en plein soleil.

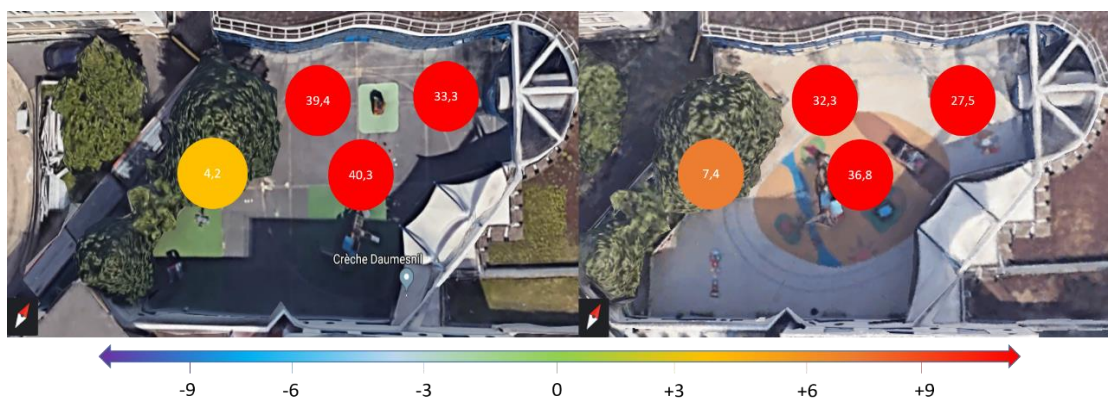

Figure 4 Cartographie de l' écart $T_{mrt,i} - T_{air,ref}$ dans la cour Daumesnil

**Conclusion**

L'impact de la transformation de la cour de l'école maternelle Daumesnil a été évaluée par deux campagnes de mesure mobile réalisées avant et après travaux consistant en la pose d'un revêtement réfléchissant et drainant, la plantation de deux nouveaux arbres et le remplacement des pieds d'arbres existants par de la terre végétale.

L'augmentation de l'albédo du revêtement de la cour permet une réduction de l'ordre de 5°C de l'écart entre la température radiante et la température ambiante, permettant ainsi de limiter l'échauffement de l'air ambiant. Toutefois, on observe l'inverse sous l'arbre. Deux facteurs ont pu contribuer à cette dégradation : une augmentation du rayonnement solaire réfléchi par le revêtement de la cour (albédo +0,2) d'une part, et un printemps post-travaux plus sec d'autre part. Ces résultats rappellent également l'importance du stress hydrique sur la performance rafraîchissante de la végétation.

D'autres campagnes dans les années à venir pourraient permettre de départager les deux phénomènes pour le site en question, notamment à l'occasion d'un printemps plus humide. De plus, la désimperméabilisation du site étant récente, les végétaux déjà présents sur place n'ont possiblement pas eu le temps de développer leur réseau racinaire pour en bénéficier pleinement.

L'impact cumulé des stratégies de rafraichissement, à savoir désimperméabilisation, végétalisation et augmentation d'albédo des surfaces ne signifie en tout cas pas forcément amélioration du stress thermique.

Etant donné que seulement deux campagnes de mesures mobiles ont été réalisées, la robustesse statistique des résultats présentés ne peut être démontrée. Par ailleurs, la station témoin utilisée ici est située à plusieurs kilomètres du site d'étude. L'étude d'un point non impactés par les travaux de réaménagement permettrait de tenir compte de la variabilité temporelle de ces mesures (Chanial et al, 2021).

Par la suite, lors de l'évaluation des 10 cours pilotes du projet FEDER UIA OASIS, nous combinerons l'utilisation des mesures mobiles à une paire de stations météo situées à moins de 500 mètres l'une de l'autre, ce qui permettra une analyse statistique fine en un point de la cour.

### Remerciements

### Bibliographie

ASTM E1918-16. 2016. "Standard Test Method for Measuring Solar Reflectance of Horizontal and Low-Sloped Surfaces in the Field." *ASTM International* i (Reapproved 2015): 18–20. https://doi.org/10.1520/E1918-16.

Błazejczyk, Krzysztof, Peter Broede, Dusan Fiala, George Havenith, Ingvar Holmér, Gerd Jendritzky, Bernhardt Kampmann, and Anna Kunert. 2010. "Principles of the New Universal Thermal Climate Index (UTCI) and Its Application to Bioclimatic Research in European Scale." *Miscellanea Geographica* 14: 91–102. https://doi.org/10.2478/mgrsd-2010-0009.

Bowler, Diana E., Lisette Buyung-Ali, Teri M. Knight, and Andrew S. Pullin. 2010. "Urban Greening to Cool Towns and Cities: A Systematic Review of the Empirical Evidence." *Landscape and Urban Planning* 97 (3): 147–55. https://doi.org/10.1016/j.landurbplan.2010.05.006.

Erell, Evyatar, David Pearlmutter, Daniel Boneh, and Pua Bar Kutiel. 2014. "Effect of High-Albedo Materials on Pedestrian Heat Stress in Urban Street Canyons." *Urban Climate* 10 (P2): 367–86. https://doi.org/10.1016/j.uclim.2013.10.005.

Hendel, Martin. 2016. "Pavement-Watering in Cities for Urban Heat Island Mitigation and Climate Change Adaptation." https://tel.archives-ouvertes.fr/tel-01258289.

Lai, Dayi, Wenyu Liu, Tingting Gan, Kuixing Liu, and Qingyan Chen. 2019. "A Review of Mitigating Strategies to Improve the Thermal Environment and Thermal Comfort in Urban Outdoor Spaces." *Science of the Total Environment*. Elsevier B.V. https://doi.org/10.1016/j.scitotenv.2019.01.062.

Leconte, François, Julien Bouyer, Rémy Claverie, and Mathieu Pétrissans. 2015. "Estimation of Spatial Air Temperature Distribution at Sub-Mesoclimatic Scale Using the LCZ Scheme and Mobile Measurements." *9th International Conference on Urban Climate & 12th Symposium on Urban Environment*, no. July: 1–12.

Mairie de Paris. 2017. *Stratégie de Résilience de Paris*. Paris, France.. 2018. *Plan Climat de Paris*. Paris, France.

Meehl, Gerald A., and Claudia Tebaldi. 2004. "More Intense, More Frequent, and Longer Lasting Heat Waves in the 21st Century." *Science* 305 (5686): 994–97. https://doi.org/10.1126/science.1098704.

Mohegh, Arash, Ronnen Levinson, Haider Taha, Haley Gilbert, Jiachen Zhang, Yun Li, Tianbo Tang, and George A. Ban-Weiss. 2018. "Observational Evidence of Neighborhood Scale Reductions in Air Temperature Associated with Increases in Roof Albedo." *Climate* 6 (4): 98. https://doi.org/10.3390/cli6040098.

Pasquill, F. 1961. "The Estimation of the Dispersion of Windborne Material." *The Meteorological Magazine* 90 (1063): 33–49.

Perkins-Kirkpatrick, S. E., and P. B. Gibson. 2017. "Changes in Regional Heatwave Characteristics as a Function of Increasing Global Temperature." *Scientific Reports* 7 (1). https://doi.org/10.1038/s41598-017-12520-2.

Tsin, Pak Keung, Anders Knudby, E. Scott Krayenhoff, Hung Chak Ho, Michael Brauer, and Sarah B. Henderson. 2016. "Microscale Mobile Monitoring of Urban Air Temperature." *Urban Climate* 18: 58–72. https://doi.org/10.1016/j.uclim.2016.10.001.

M.Chanial, G. Karam, S.Parison, M.Hendel, et L.Royon 2021." Cartographie du stress thermique au sein d'une cours d'école parisienne : Couplage de mesures microclimatiques fixes et mobiles" *34ème colloque international annuel de l'AIC*